\documentclass[aps,prb,groupedaddress,twocolumn,10pt]{revtex4-1}

\begin{document}

\setcitestyle{numbers}
\setcitestyle{square}

\newcommand{\be}{\begin{equation}}
\newcommand{\ee}{\end{equation}}
\newcommand{\bea}{\begin{eqnarray}}
\newcommand{\eea}{\end{eqnarray}}

\title{Reckoning with the Mother of all non-Fermi liquids:\\
 alien bosonization vs predator holography}

\author{D. V. Khveshchenko} 
\affiliation{Department of Physics and Astronomy, 
University of North Carolina, Chapel Hill, NC 27599}

\begin{abstract}
\noindent
This note addresses 
the problem of computing fermion propagators in a broad variety of strongly correlated systems that can be mapped onto the theory of fermions coupled to an (over)damped bosonic mode. A number of the previously applied approaches and their results are reviewed, including the conventional diagrammatic resummation and eikonal technique, as well
as the 'experimental' higher dimensional bosonization and generalized  
(i.e., 'bottom-up' or 'non-AdS/non-CFT') holographic conjecture. It appears that, by and large, those results remain either in conflict or incomplete, thereby suggesting that the ultimate solution to this ubiquitous problem is yet to be found. 
\end{abstract}
\maketitle

{\it Prologue}\\

\noindent
A quest into the various metallic states of interacting fermions has been continuing over the past few decades, its main goal being a systematic classification of those compressible 
states and their properties. 
Despite all the effort, though, the only fully understood is the 
classical Fermi liquid (FL) while (possibly countless) 
deviations from it remain largely unexplored and still need to be systematized.

Much of the previous studies revolved around a broad class of systems governed by some long-ranged and/or retarded two-fermion interactions that are often associated with ground state instabilities and concomitant non-Fermi-liquid (NFL) behaviors which may occur even in those systems whose microscopic Hamiltonians involve only short-range couplings.

In the close proximity to a quantum phase transition, an effective singular coupling can be mediated by (nearly gapless) collective excitations of an emergent order parameter of charge, spin, or other nature. Important examples include, both, the physical and effective finite-density QED, quark-gluon plasma in QCD, (anti)ferro-magnetic fluctuations in hole-doped cuprates and heavy fermion materials, Ising nematic and other Pomeranchuk/Lifshitz-type transitions of the 
itinerant Fermi surfaces (FS), even-denominator compressible Quantum Hall Effect (QHE), etc. 

The quantum theory of strongly correlated fermions has long been in 
a strong need of non-perturbative techniques the use of 
which would allow one to proceed beyond the customary (yet, often 
uncontrollable in the regime of interest) approximations 
when analyzing generic (non-integrable) systems.

In that regard, it's been claimed that a possible way out of the lingering stalemate
can be found along the lines of the once 
popular, then (nearly) abandoned, and recently resurrected idea of higher-dimensional bosonization 
or provided by the never proven, yet massively entertained, conjecture of 
the generalized ('non-AdS/non-CFT') holographic duality.

Given the current interest in such 'experimental'  
techniques and for the sake of elucidating their true status 
it would be worth comparing their predictions, as well as contrasting them 
against the other available results.\\

{\it Diagrammatic(s) Rules}\\

\noindent
In many of the diverse reincarnations of the problem of finite density fermions 
with the interactions mediated by gapless bosonic excitations 
the propagator of the latter 
conforms to the general expression
\be
D(\omega,{\bf k})={1\over |\omega|/{ k}^\xi+{k}^\rho}
\ee
Hereafter, Matsubara frequencies will be used
while all the dimensionful factors will be absorbed into the proper units of momentum ($k_F$) and energy
($ g^{2/1-\eta}(v_Fk_F)^{(1+\eta)/\eta-1}/N$, see Eq.(4) below)
 where $k_F$ and $v_F$ are the Fermi momentum and velocity,
$g$ is the coupling strength, and $N$ is the number of fermion species.
Likewise, all the inessential numerical factors will be dropped. 

Among the practically important examples of this 'Mother of all NFLs' 
are such seemingly disjoint topics as electromagnetic (i.e., Abelian gauge field) 
skin effect in metals \cite{reizer} and  quark-gluon (non-Abelian) plasmas \cite{qcd}, 
spin \cite{palee} and charge \cite{walter1} fluctuations in itinerant ferromagnets and Ising quantum nematics, as well as compressible Quantum Hall effect with screened 
repulsive interactions \cite{qhe}, in all of which situations $\xi=1$ and $\rho=2$. 
By contrast, normal skin effect and antiferromagnetic fluctuations in doped Mott insulators  
are described by $\xi=0, \rho=2$, while compressible Quantum Hall Effect with the 
unscreened Coulomb interactions corresponds to $\xi=1, \rho=1$.

Over several decades much effort has been made 
towards ascertaining the effects of the interaction (1) on the FL 
propagator with a finite chemical potential $\mu$ 
\be
G_{0}(\omega,{\bf k})={1\over {i\omega-\epsilon_{\bf k}+\mu}}
\ee
whose Fourier transform in the spacetime domain reads  
\bea
G_0(\tau, {\bf r})=
\int d{\bf k}d\omega e^{i{\bf k}{\bf r}-i\omega\tau}G_0(\omega,{\bf k})\nonumber\\
=\sum_{\pm}{e^{\pm ik_Fr}(r\pm i\tau)\over r^{1/2}(\tau^2+r^2)}
\eea
Previous diagrammatic approaches to this problem sought out to investigate the 
stability of the first-order self-energy  
\be
\Sigma(\omega)\sim \omega^{\eta},~~~~~~~~\eta={d-1+\xi\over \xi+\rho}
\ee
against higher-order corrections. For a choice of parameters conspiring to yield $\eta=1$
the self-energy (4) acquires an extra factor $\ln\omega$. 

In the early analyses it was  
argued that the self-energy retains its functional form (4) to all orders in perturbation theory for any finite $N$, provided that the FS curvature is properly accounted for \cite{chub,chubdvk,chubsed}. 
This conclusion was drawn on the basis of 
self-consistent Eliashberg-type diagrammatics which, in turn, relies on 
the generalized Migdal theorem to control vertex corrections. 

Utilizing the conjectured all-orders result (4) one arrives at the expression 
\be
G_{diag}(\omega,{\bf k})={1\over i\omega-\epsilon_k+\mu+\Sigma(\omega)}
\ee
where the self-energy is a power-law function of energy with only a weak momentum dependence.   
To account for a FS curvature $\kappa$
the fermion dispersion 
can be expanded in the vicinity of the (Luttinger) FS traced by the unit normal $\bf n$ 
\be
\epsilon_k=\mu+v_F({\bf k}{\bf n}-k_F)+\kappa({\bf k}\times{\bf n})^2
\ee
Upon Fourier transforming (5) one finds that at the largest  
spatial separations the equal-time propagator demonstrates a power-law
behavior 
\be 
G_{diag}(\tau\to 0,{\bf r})\sim {e^{\pm ik_Fr}\over r^{3/2+\eta_1}}       
\ee
which decays with the distance faster than its non-interacting counterpart (3) for 
any $\eta>0$. In the self-consistent Eliashberg theory, 
the anomalous exponent was found to take the universal value $\eta_1=1/2$ \cite{chubdvk,chubsed}. 
By contrast, the alternate approaches found 
different values of $\eta_1$, as well as a  
faster-than-algebraic behavior at intermediate distances, $N\lesssim r\lesssim 1/N^2\kappa^3$,
which range expands for $N\to 0$ and/or $\kappa\to 0$ (see next sections).
 
Moreover, in the complementary limit of large temporal separations the leading term in the all-orders 
'near-field' propagator retains its non-interacting form \cite{chubdvk} 
\be 
G_{diag}(\tau,{\bf r}\to 0)\approx G_{0}(\tau,{\bf r}\to 0)
\sim {e^{\pm ik_Fr}\over \tau r^{1/2}}                
\ee 
although the sub-leading corrections bear some non-trivial $\tau$-dependence 
(see next sections).

It has also noted that the algebraic behaviors (7,8) hold due to the presence of a pole in the integration over $\epsilon_k$ in (5) while in its absence a different functional behavior sets in. 
However, the latter was predicted to occur only in the (arguably, unphysical)
limit $N\to 0$.

Contrary to the earlier expectations, though, the refined analyses 
of higher-order corrections to (4) 
found them to be singular, albeit suppressed by extra powers of $1/N$ \cite{sslee}. 
A number of 
attempts to get an analytic handle on the higher-order effects has been made \cite{mross}
but their full bearing on the problem of interest remains unclear.

Furthermore, a naive generalization of the above calculations to finite temperatures 
appears to be problematic as (4) picks up a singular contribution $\Sigma(0)$ \cite{walter1}.
This problem is particularly severe in those situations where gauge invariance prevents 
the mediating transverse gauge field ${\bf A}_{\perp}={\bf A}\times{\bf k}/k$  
from developing a thermal mass (no magnetostatic screening in normal metals).

Interestingly, one encounters the same formal hurdle in the problem of a particle 
moving in random magnetic field (RMF), the random field's variance substituting for 
the temperature. Since scattering off of static ($\omega=0$) magnetic field disorder correlated as
$<{\bf A}_{\bf k}{\bf A}_{\bf -k}>\sim T/{k}^{\rho}$ is purely elastic  
the problem has a purely single-particle nature. Yet, for $\rho\geq d$ 
the standard self-consistent Born approximation gets invalidated due to the infrared divergence 
at small momenta.

Note, in passing, that absolute majority of the cited work on the RMF problem in $d=2$  
assumes short-range correlations, $<{\bf A}_{\bf k}{\bf A}_{\bf -k}>\sim const$, which allow  
for applications of the powerful machinery of $2d$ conformal field theory (CFT). 
However, a non-perturbative solution to the 
long range-correlated problem is available, too \cite{aim1,rmf}.      
Properly adapted, if offers a workaround for the technical problem posed by the singular zero-frequency harmonic at finite $T$.

A simpler - yet, questionable - practical recipe for dealing with this harmonic would be to ignore it altogether - as an artifact of the gauge-non-invariant nature of the fermion propagator- or, more formally, have it absorbed into the renormalized chemical potential.

Alternatively, in those situations where the issue of gauge invariance is moot (e.g., fluctuations of electrostatic potential in metals), an emergent thermal (Debye) mass $m_{th}(T)\sim(T\ln T)^{1/2}$
of the bosonic mediator provides for the necessary regularization \cite{walter1}.  
\\ 

{\it Straight-forward Eikonal}\\ 

\noindent
Another approach to the problem of fermions with the interactions (1)
was proposed in the small-$N$ limit where there is no fermion back-reaction
as any fermionic loops are suppressed.
This limit was first addressed in the context of summing over the so-called 
'maximally crossed' diagrams \cite{stamp}. 
However, shortly after its first applications it was realized that this method represents a simplified form 
of the general eikonal approach \cite{stampdvk}. 

The most straightforward (in the geometrical sense as well) 
eikonal calculation focuses exclusively on the Debye-Waller 'damping factor' picked up by the fermion's wave function while the latter is moving down its straight-line semiclassical path.   
Computing the average $<\exp(i\int d{\bf r}{\bf A})>$ 
over the bosonic fluctuations with the interaction function
$D(\omega,{\bf k})=<{\bf A}{\bf A}>$ given by (1)  
one obtains 
\be
G_{eik}(\tau,{\bf r})=G_0(\tau,{\bf r})e^{-S_{eik}(\tau,{\bf r})}
\ee
where the exponential 'damping factor' reads 
\bea
S_{eik}(\tau,{\bf r})=\int d\omega d^d{\bf k}
D({\omega,{\bf k}})G(\omega,{\bf k})G_0(-\omega,-{\bf k})\nonumber\\
{(1-\cos(\omega\tau-{\bf k}{\bf r}))}~~~~~~~~~~~~
\eea
First, neglecting the FS curvature one readily finds the lowest-order eikonal action 
\be
S_{eik}(\tau,{\bf r})\sim {r\over |r-i\tau|^{\eta}}
\ee
whereas for $\eta=1$ 
the frequency integration produces an additional factor $\sim\ln|r-i\tau|$. 

For example, in $d=2$ and for $\eta=2/3$ (11) gives rise to the asymptotics 
\be 
G_{eik}(\tau\to 0,r)\sim {1\over r^{3/2}}\exp(-{r^{1/3}})
\ee
and
\be 
G_{eik}(\tau,r\to 0)\sim {1\over \tau r^{1/2}}\exp(-{r/\tau^{2/3}}) 
\ee
which were obtained in \cite{chubdvk,lawler,nordita12,banks}, barring minor differences (see next section).
The expressions (12,13) yield the 'off-FS' Fourier transform 
\be
G_{eik}(\omega,{\bf k})={1\over i\omega-v_F(k-k_F)}
\exp(-{\omega\over |k-k_F|^{3/2}})
\ee
that was first found in \cite{aim2} and recently re-discovered in \cite{nordita12,banks}.
Its complementary 'on-FS' behavior, $G_{eik}(\omega,k_F)\sim\exp(-1/\omega^{1/2})$, appeared in \cite{stampdvk}.  

However, as pointed out in \cite{aim2} and reiterated in \cite{chubdvk,chubsed}, 
for a finite $\kappa$ the faster-than-algebraic
decay (14) emerges only in the limit $N\to 0$ where  
the pole in (5) as a function of $\epsilon_k$ disappears.
Furthermore, if the entire $\omega$-linear term (Landau damping)
in the interaction function (1) were generated in the one-loop approximation 
then the behavior (14) could only be attained by taking  
the double limit $N\to 0$, $k_F\to\infty$ with $Nk_F$ kept constant
in order to suppress all the higher-order fermion loops \cite{nordita12}. 

The eikonal approximation can be systematically improved and also extended to the two- (and, potentially, any-) particle amplitudes \cite{stampdvk,aim2,nordita12}. In particular,  
the two-particle (charge/spin) density correlation functions in $2d$ 
conform to the general expression
\be
<\rho(\tau,r)\rho(0,0)>= {a_0\over r^3}+{b_0\over \tau^2r}+
{\sin(2k_Fr)\over r^{1-\eta_2}}
({a_{2k_F}\over r^{2}}+{b_{2k_F}\over \tau^2})
\ee
Notably, at low momenta its Fourier transform retains its free form 
$<\rho(\omega,{\bf k})\rho(-\omega,-{\bf k})>\sim const+\theta(v_Fk-\omega){\omega/k}$
while at momenta close to $2k_F$ the density response 
\be
<\rho(\omega,{\bf k})\rho(-\omega,-{\bf k})>\sim {(a|k-2k_F|+b|\omega|^{2/3})^{1-\eta_2}}
\ee
demonstrates a cusp or even a true singularity, depending on the anomalous exponent $\eta_2$ 
(at large $N$ one finds $\eta_2\sim 1/N$, though) \cite{aim2,stampdvk,nordita12}.\\

{\it Legacy Bosonization}\\

\noindent
Developing an approach to generic condensed matter problems that would be 
as powerful and versatile as $1d$ bosonization has long remained as desirable, as it has been elusive. 
The vigorous early studies in that direction \cite{bos,bosrice,bosdvk} were followed by a long period of oblivion, although there has 
been some revival of interest, as of lately \cite{son}. 
Thus far, however, concrete predictions of this technique
amounted to reproducing the classic FL results \cite{bos} or, else,
delivering some novel ones which appear to be quite 
similar to those obtained by means of the eikonal approach.

At the practical level, the simplest version of the 
bosonization technique is described  
as a triangulation procedure in the course of which the FS gets divided onto a large number of 
'patches' whose size decreases upon renormalization.
The idea is that a predominantly forward scattering makes physics quasi-$1d$ and subject to the similar (albeit, approximate in $d>1$) Ward identities \cite{walter2}.
One might then hope that this would suffice for permitting the use of the $1d$ rules of substituting 
bosonic ('plasmon') density modes for fermion bilinears at small energies/momenta. 

Conceivably, this condition could indeed be met in a situation where an auxiliary 
patch size $\Lambda\ll k_F$ drops out of the calculation due to the singular nature of the interaction. 
For the interaction (1)
this happens when the integral $\int d^{d-1}{\bf k}D(\omega,{\bf k})$ over the momenta
tangential to the FS converges
(i.e., for $d-1<\rho$). It might still be marginally acceptable 
when the resulting dependence is logarithmic but would be
harder to justify if the above integral diverged at the upper limit set by $\Lambda$. 

The equivalent bosonic action is formulated 
in terms of the phase variable $\phi(\tau,{\bf r},{\bf n})$ labeled by its 
location on the surface of constant Fermi energy traced by the  
unit normal ${\bf n}$ (or the corresponding $d-1$ angular parameters)  
\bea
S_{bos}=\int d{\bf r}d\tau 
\sum_{\bf n}
(\partial_{\tau}\phi({\bf n}{\bf \nabla}\phi)-{v_F}({\bf \nabla}\phi)^2)~~~~~~~~~~~\\
+{1\over 2}\int d{\bf r}_1d\tau_1 d{\bf r}_2d\tau_2
\sum_{{\bf n}_1,{\bf n}_2}
D(\tau_{12},{\bf r}_{12})
({\bf n}_1{\bf \nabla}\phi_1)({\bf n}_2{\bf \nabla}\phi_2)\nonumber
\eea
where $D(\tau,{\bf r})$ is the Fourier transform of (1) 
and the sum is over the FS patches of a progressively decreasing size.

The proposed 'ad hoc' construction of the single-particle operator \cite{bos}
\be
\psi_{bos}(\tau,{\bf r})
=\sum_{\bf n}e^{i{\bf k}_F{\bf r}+i\phi(\tau,{\bf r},{\bf n})}
\ee
ignores any operator-valued factors $a$ $la$ $Klein$ that in $1d$ make the thus-obtained fermion 
operators obey the standard commutation relations at different FS points. 
Therefore, while adequately reproducing the low-energy/momentum FS hydrodynamics  
at the level of (charge/spin)density response functions 
the bosonization 
recipe (18) apriori may not be sufficient when it comes to single-particle amplitudes.

Nonetheless, 
the single-particle propagator obtained with the use of (18) is given by the expression 
\bea
G_{bos}(\tau, {\bf r})=\sum_{\bf n}
e^{i{\bf k}_F{\bf r}-{1\over 2}<(\phi(\tau,{\bf r},{\bf n})-
\phi(0,{\bf 0},{\bf n}))^2>}
\eea
where the brackets stand for the average over the phase fluctuations  
governed by the Gaussian action (17).

In the non-interacting case the corresponding integral is logarithmic, 
thus resulting in the angular expansion of the free propagator (3) with the 
dispersion linearized near the FS
\bea
G_0(\tau, {\bf r})=\sum_{\bf n}\int d\omega d{\bf k}
{e^{i({\bf n}k_F+{\bf k}){\bf r}-i\omega\tau}\over i\omega-{v}_F{\bf n}{\bf k}}
\delta^{(d-1)}({\bf k}\times{\bf n})\nonumber\\
=\sum_{\bf n}
{e^{ik_F{\bf n}{\bf r}}\over {\bf n}{\bf r}-i\tau v_F}~~~~~~~~~
\eea
Neglecting any effects of the FS curvature one obtains a formula  
reminiscent of the eikonal propagator (9)  
\be 
G_{bos}(\tau, r)\sim{1\over r-i\tau}\exp(-S_{eik}(\tau,r))
\ee
governed by the action (11). 
It was then argued in \cite{chubdvk,lawler,kopietz} that 
the asymptotic $S_{eik}(\tau\to 0, r)\sim r^{1/3}$ may only hold at intermediate spatial distances.
For any finite $N$ the FS curvature is generally expected to alter this behavior  
at the longest distances, $r\gtrsim 1/N^2\kappa^3$, 
thereby resulting in such conflicting predictions as that of    
no significant effect at all \cite{lawler} to restoring the standard FL behavior \cite{kopietz}, to
yielding an anomalous power-law decay (7) with $\eta_1\sim 1/N$ \cite{chubdvk}).

Also, with the FS curvature taken into account 
the complementary large-$\tau$ asymptotic $S_{eik}(\tau, r\to 0)\sim r/\tau^{2/3}$  
was predicted to turn into $S_{eik}(\tau,0)\sim (\ln\tau)/\tau^{2/3}$ \cite{lawler} or 
$S_{eik}(\tau,0)\sim (\ln\tau)/\tau$ \cite{chubdvk}. The marked difference between these predictions
takes its origin in the different ways of handling the near-double pole in (10)
which is only split at a finite $\kappa$. 
 
Although the original bosonization approach appears to be rather similar to the eikonal 
technique, it can be further improved.
To that end, its more sophisticated version (see \cite{bosdvk} 
and its recent redux \cite{son})
can be formulated in terms of a path integral over
the Boltzmann distribution function which plays the role of a collective bosonic field variable. The rigorous formulation of this approach makes use of the Costant-Kirillov method of 
quantization on the coajoint 
orbits of the phase space volume-preserving diffeomorphisms   
whose generators obey $W_{\infty}$ algebra.
 
However, while potentially 
providing a systematic way to refine (21) by 
taking into account the non-linear terms (higher derivatives of $\phi(\tau,{\bf r},{\bf n})$)
in the effective hydrodynamics, this technique has not yet demonstrated its full potential.
Same can be said about a very different scheme (which is formally exact by design, too) 
that exploits intrinsic supersymmetry by introducing ancillary ghost fermions \cite{efetov}.

Albeit still waiting to be explored, neither way of improving on the lowest-order eikonal/bosonization results is expected to be an easy task. In that regard, the 
recent (largely, verbal) claim  \cite{banks} of exactness of the asymptotic (21)  
seems much too simple to be true.\\
 
{\it Wonton Holography}\\

\noindent
Compared to what it has been just recently \cite{hol}, the seemingly endless flurry of  
holographic publications in JHEP, PRD, and other 
traditional 'condensed matter oriented' venues has been 
steadily coming to a mere trickle. 
Those few holographic exercises that do occasionally pop out 
still tend to begin with the mantra  'holography is well known to be an established method for studying strongly correlated systems'. However, this optimistic reassurance
often appears to be in a rather stark contrast with the typical summary 
that sounds more like 'as no unambiguous agreement with 
experiment was found, the problem is left to future work'.

Also, much of the original thrust towards boldly treating an arbitrary condensed matter system 
of interest as yet another application of some opportunistically chosen weakly-coupled semiclassical gravity    
has retreated into a 'safer-haven' topic of hydrodynamics (which, while highlighted and revitalized by holography, can be - and of course had long been - successfully discussed without ever mentioning the latter).  

On the outside, it may seem as though the heuristic 'holo-hacking' 
(a.k.a. 'bottom up' or 'non-AdS/non-CFT') approach 
tends to pick out its favorite gravity-like bulk theory
on the basis of such physically compelling reasons as an existence of the previously 
found classical solutions and normal modes' spectra, availability of the numerical simulation 
software, or mere need to engage students with the tangible computational tasks.

However, apart from having become a massive and customary practice,
there hasn't been much effort made towards any serious justification of 
neither the overall holographic scheme, 
nor its specific 'dictionary' which was copy-pasted 
from the original string-theoretical framework.
In that regard, it might be worth keeping in mind that just because 
everyone else on a highway may be driving above the posted speed limit does not by itself make it legal. 

In light of the above, comparing holographic propagators
to the predictions of other techniques could provide an additional  
testing ground for, both, the alternate methods as well as the holographic approach itself.

Over the past $15$ years, the most common holographic workhorse has been  
the Einstein-Maxwell-dilaton (EMD) theory described by the $d+2$-dimensional 
Lagrangian for the bulk metric, gauge, and scalar (dilaton) fields \cite{hol}
\begin{equation}
S_{EMD}=\int{1\over 2}(\partial\varphi)^2-R-{d(d+1)\over L^2}
e^{\delta\varphi}+{1\over 2}e^{\zeta\varphi}F_{\mu\nu}^2
\end{equation}
where $\int=\int d\tau du d^d{\bf r}{\sqrt g}$ stands for a covariant $d+2$-dimensional volume integral
and $R$ is a scalar curvature. The third term represents 
a (negative) cosmological 'constant' of the asymptotically 
anti-de-Sitter space $AdS_{d+2}$ of radius $L$ and metric determinant $g$.

The vacuum solutions of (22) include 
a broad class of diagonal (Euclidean) metrics 
\be
ds^2=f(u)d\tau^2+g(u){du^2}+h(u)d{\bf r}^2
\ee
with the algebraically decaying components
\begin{equation}
g(u)=(L/u)^{2\alpha},~~h(u)=(L/u)^{2\beta},~~f(u)=(L/u)^{2\gamma}
\end{equation}
where the exponents are functions of $\delta$ and $\zeta$.
These metrics can not be readily extended to the asymptotic ultraviolet regime 
and, therefore, need to be regularized in the near-boundary limit ($u\to 0$)
where they revert back to the $AdS_{d+2}$ background.

Among all the different parametrizations of the 'radial' variable $u$   
a particularly convenient is the one with only two essential parameters:
dynamical exponent $z$ and hyperscaling violation (HV) parameter $\theta$ 
\be 
z={1+\gamma-\alpha\over 1-\alpha+\beta},~~~~\theta=d{1-\alpha\over 1-\alpha+\beta}
\ee
in terms of which the metric 
(23) takes the standard HV form 
\be
ds^2=u^{2\theta/d}({d\tau^2\over u^{2z}}+{du^2+d{\bf r}^2\over u^2})
\ee 
The indexes $z$ and $\theta$    
manifest themselves through the scaling of the boundary 
excitation spectrum, $\omega\to\omega/ \lambda^z$, 
and the interval, $ds\to\lambda^{\theta/d} ds$ under the change of  momenta ${\bf k}\to{\bf k}/\lambda$.  

Among all the classical solutions of the theory (22) there is a special class 
of Lifshitz metrics ($\theta=0$) which were discovered in  
the semiclassical (Thomas-Fermi) analysis of matter back-reaction on the metric, as well as 
in the 'electron star' scenarios, etc. \cite{lif}.
 
More generally, any viable solutions of (22) must obey certain stability 
('null energy') conditions \cite{hol} 
\be 
(d-\theta)(d(z-1)-\theta)\geq 0,~~~(z-1)(d+z-\theta)\geq 0
\ee

{\it Holographic Propagators}\\

\noindent
The early holographic studies of fermion propagators \cite{holferm} 
produced a number of intriguing results, 
including multiple Fermi surfaces (which merge into one critical 
'Fermi ball' in some extreme limits), dispersionless poles, and
oscillatory frequency dependence (which was later shown not to arise in more 
systematic 'top down' constructions \cite{hol}), etc. 
A physical interpretation of those results is impeded by the fact that much of this work is numerical.

A simple and amenable to analytical treatment semiclassical calculation can be performed  
in the regime $mL\gg 1$ where $m$ is a mass of the conjectured dual bulk fermion \cite{holferm,hv}.
In this regime, the fermion's paths contributing to various 
quantum-mechanical amplitudes follow closely the classical boundary-to-boundary 
trajectories (geodesics) derived from the (imaginary-time) action
\be
S_{hol}=m\int du {\sqrt {g(u)+f(u)({d\tau\over du})^2+h(u)({d{\bf r}\over du})^2}} 
\ee 
by varying over $\tau(u)$ and ${\bf r}(u)$.

Evaluating this action on its geodesic one obtains   
\be
S_{hol}(\tau, {\bf r})=m\int^{u_t}_0 du{\sqrt {g(u)\over q(u)}}
\ee
Here $q(u)=1-{\pi}^2_r/h(u_t)-\pi^2_\tau/f(u_t)$  
is a function of the conjugate momenta ${\pi}_r=\delta S/\delta (d{\bf r}/du)$ 
and $\pi_\tau=\delta S/\delta (d\tau/du)$ obeying the equations
\be
{r}={\pi}_r
\int^{u_t}_0 {du\over h(u)}{\sqrt {g(u)\over q(u)}},~~~
\tau=\pi_\tau
\int^{u_t}_0 {du\over f(u)}{\sqrt {g(u)\over q(u)}}
\ee
where the upper limit $u_t$ is a turning point 
given by the condition ${q}(u_t)=0$.

While an explicit analytic computation of (29) can only be performed in 
some special cases, the one-parameter space/time dependencies  
can be readily found for a broad variety of metrics.
Specifically, for the HV metric (26) one obtains \cite{hv,holdvk1} 
\be
S_{hol}(\tau\to 0,r)\sim r^{\theta/d},~~~
S_{hol}(\tau,r\to 0)\sim \tau^{\theta/dz}
\ee
Notably, in the absence of hyperscaling violation ($\theta=0$)
both these asymptotics become either constant (less likely) or logarithmic (more likely, see below). 
Thus, if the classical EMD Lagrangian (22) were to represent 
a valid bulk dual of a boundary theory with the gauge-like interaction (1) the asymptotics (31)  
would not be readily reconcilable with the eikonal/bosonization results (11,21)
which depend primarily on $z$ (via $\eta$) rather than $\theta$.

Specifically, matching the large-$r$ behaviors produced 
by the eikonal/bosonization  and holographic calculations requires 
$z=({1-\theta/d})^{-1}$, but even under this condition
the results pertaining to the complimentary long-$\tau$ regime 
would still remain in conflict with (11,21). 

Alternatively, the semiclassical analysis can be applied 
directly to the bulk wave function in the energy-momentum representation 
$\psi(u,\omega,{\bf k})$. It describes a bulk fermion 
subject to the effective single-particle potential in the radial holographic direction \cite{holferm}
\be
V(u)=m^2+{{\bf k}^2\over h(u)}+{\omega^2\over f(u)}
\ee
and is composed of the two independent solutions which read 
\be
\psi_{\pm}(u,\omega,{\bf k})\sim {1\over V^{1/4}(u)}
e^{\pm i\int^{u_t}_udu^{\prime}{\sqrt {g(u^{\prime})V(u^{\prime})}}}
\ee
in the classically permitted region $0<u<u_t\sim(k/\omega)^{1/\gamma-\beta}$.

Imposing the proper boundary conditions and following the holographic dictionary \cite{hol} 
one then defines the propagator as a reflection coefficient for the 
wave incident at the boundary
\be
G_{hol}(\omega,{\bf k})=
{a_+(u,\omega,{\bf k})+e^{-S_{hol}}b_+(u,\omega,{\bf k})
\over 
{a_-(u,\omega,{\bf k})+e^{-S_{hol}}b_-(u,\omega,{\bf k})}}
|_{u\to 0}
\ee
where the coefficients $a_{\pm}, b_{\pm}$ 
pertain to the large/small-$u$ asymptotics of the solutions $\psi_{\pm}$ 
which have to be matched in the
overlapping intermediate regime.

Expanding (34) near the pole under the assumption of no accidental 
pole-skipping and comparing with (5) one identifies a pertinent self-energy $\Sigma(\omega,{\bf k})\sim e^{-S_{hol}(\omega,{\bf k})}$ determined by 
the tunneling action through the classically forbidden radial region \cite{holferm,holdvk1}
\be
S_{hol}(\omega,{\bf k})\sim ({{k}^{1+\gamma-\alpha}\over \omega^{1+\beta-\alpha}})^{1\over \gamma-\beta}
\ee
which shows exponential dependence on, both, energy and momentum.
Interestingly, for a HV metric (26) it becomes  $S_{hol}(\omega,{\bf k})\sim ({{k}^{z}/\omega})^{1\over z-1}$ and depends solely on $z$ but not $\theta$. In that regard the Fourier transform of 
the exponential of (35) might have, in principal, 
a better chance to agree with (11) - at least, for certain specific values of $z$. 

A different behavior (unattainable in the case of a HV metric (26) with 
finite $z$ and $\theta$) occurs for $\alpha=\beta+1$ in which case 
the integral in (33) diverges at $u\to 0$. This peculiar NFL regime, 
dubbed 'local criticality', is characterized by the propagator 
\be
G_{loc}(\omega,{\bf k})={1\over a(k)+b(k)\omega^{\nu(k)}}
\ee
where $a(k),b(k)$, and $\nu(k)\sim k$ are non-singular 
functions of momentum that can, in general, produce multiple poles identified as 
the distinct ('fractionalized') FS \cite{holferm}.

Fourier transforming (36)
is complicated by the fact that $G(\omega,{\bf k})$ is not
analytically known across the entire range of its arguments.
However, the fast (and/or furious) Fourier transformation via a saddle point suggests  
the following form of this function in the spacetime domain 
\be
G_{loc}(\tau,{\bf r})\sim\exp(-{\sqrt {({k_Fr})^2+(\nu_F\ln\tau)^2}})
\ee
where $\nu_F=\nu(k_F)$. This function 
rapidly decays at finite spatial distances (different 'lattice sites') 
while demonstrating the power-law 'on-site' self-correlations. It then describes a
system which effectively splits onto spatially uncorrelated  
'quantum impurities' each of which exhibits an effectively 
zero-dimensional ($d=0$) quantum-critical behavior.

Lately, such ultra-local scenario received much attention in the context
of the Sachdev-Ye-Kitaev (SYK)-type models \cite{syk1} 
whose conjectured dual bulk geometry $AdS_2\otimes R^d$ provides
a common near-horizon behavior of generic near-extremal black holes.

Further extensions of the original spaceless SYK behavior have been explored
by introducing a lattice of the individual SYK 'quantum dots', hybridizing the localized SYK (Majorana or Dirac) fermions with some itinerant ones, or making the SYK correlations distance-(and/or time-) dependent \cite{sykdvk1}. 

The zeroth-order 'on-site' propagator $G_0(\tau_1,\tau_2;{\bf r}_1,{\bf r}_2)=
G_{syk}(\tau_{12})\delta({\bf r}_{12})$ where $G_{syk}(\tau)\sim 1/\tau^{\nu}$ then gets modified and 
assumes the form (36) with a universal exponent $\nu=2/q$ where $q$ an even integer 
\cite{syk1}. In a generalized model which combines terms with different values of $q$ the   
one with the smallest $q$ dominates, thus yielding $\nu=1/2$ for a generic (chaotic, yet solvable) NFL state
- with the exception of a non-chaotic FL with $\nu=1$ corresponding to $q=2$.

Adding to the intrigue, there are some recent Monte Carlo results on the 
$2d$ Hubbard and $t-J$ models that have long been 
thought to represent the prototypical NFL normal state in the cuprates.
These results do not readily conform to a momentum-independent, yet strongly energy-dependent, self-energy function, showing less of energy/temperature dependence than any of the above expressions \cite{mc}.
It remains to be seen as to what this might imply for the general applicability of the theories  of fermions ('spinons') governed by the interactions (1) to the analysis of those 
microscopic models.\\ 

{\it Strange Cuprates}\\

\noindent
The theories of both, finite- and zero-density, spinons have been extensively discussed in the context of the 'strange metal' phase in the underdoped cuprates and other (arguably, even stranger) heavy-fermion compounds long before the advent of holography \cite{palee}. Once there, the applied holography quickly joined the quest into the properties of this phase that 
had long evaded a consistent and satisfactory explanation.   

Instead of going after the NFL fermion propagator, however, many of the holographic proposals 
focused on reproducing the experimental 
data in the cuprates - and often times even claimed achieving a quantitative agreement. 

In light of its intrinsically unsettled status one would have thought that it might be rather detrimental for any speculative approach to seek out not a mere qualitative but an actual quantitative, down to the number, agreement between its specific predictions and 
some pre-selected sets of experimental data. In fact, if such a quantitative agreement were indeed achieved one would have even more explaining to do (first and foremost, as to why an apriori approximate approach appears to be so unexpectedly accurate?). 

The earlier discussion of some of the popular evidence 
in support of condensed matter holography as well as the debunking of a number of its specific predictions \cite{hol} can be found in \cite{holdvk2}. However, the admirable persistence with which those predictions continued to be regularly cited in the subsequent holographic 
literature \cite{holexp} suggests that the comments of \cite{holdvk2} might have had 
been (most regretfully) overlooked. 

In fact, there is more than a single reason for which semiclassical holography (or its improvement at the level of accounting for the matter back-reaction in the Hartree-Fock approximation) - thus far, the only practical way of performing the holographic calculations \cite{hol,lif,holferm,hv} - would not have been expected to provide any quantitatively accurate results in the first place.   
There are, of course, such obvious differences from the string-theoretical holographic constructions as a low physical value of $N$ (which, in practice, often amounts to 'spin up/down') and the lack of Lorentz, translational, and/or rotational (as well as any super-)symmetries. 

Arguably, though, the most important is the fact that much of the condensed matter physics operates in the intermediate - as opposed to ultra-strong - interaction regime, while it is only the latter that is supposed to have a weakly coupled gravity as its bulk dual \cite{hol}.
Indeed, most solids form under the condition that its potential (interaction)
and kinetic energies on average balance each other out.  
This suggests that the 'bona fide' strong-coupling regime could only become attainable in some sort of a 'flat band' scenario where kinetic energy is completely quenched or, at least, significantly diminished.

In light of that, it is unsurprising that much of the recent effort towards implementing such mechanism has been centered on the SYK model and its variants \cite{syk1}
whose 'flat band' nature facilitates the existence of a holographic dual. A viable candidate to this role was proposed in the form of the  
Jackiw-Teitelboim (JT) dilaton-enhanced $1+1$-dimensional gravity \cite{syk1}. 

It is worth pointing out, though, that 
at the practical level all the holographic matching between the SYK and JT theories 
has been, so far, established within their 
low-energy sectors that are both controlled 
by a single soft Schwarzian mode ('boundary graviton'). 
So as far as the low-energy properties of the two models are concerned,  
they both allow for the same (effectively $0+1$-dimensional) description
in terms of either a fluctuating $1d$ boundary or Liouvillian-type large-$N$ matrix quantum mechanics \cite{syk1,sykdvk2}. This is not surprising given the 
intrinsically non-dynamical nature of $2d$ (and $3d$) pure gravity.   
Such a caveat notwithstanding, the low-energy SYK-JT equivalence has been repeatedly and staunchly referred to as a genuine example of holographic correspondence between the $1+1$-dimensional bulk and $0+1$-dimensional boundary theories \cite{syk1}.

As to the general HV models (22) and corresponding vacuum metrics (26), the standard list of  observables to be matched includes temperature-dependent specific heat  
\be
C(T)\sim T^{(d-\theta)/z}
\ee
and frequency-dependent optical conductivity 
\be 
\sigma(\omega)\sim \omega^{(d-2-\theta)/z}
\ee
determined by the bare scaling dimensions. 

In the underdoped cuprates, much attention has been paid to the experimentally observed 
mid-infrared algebraic behavior $\sigma\sim\omega^{-2/3}$ \cite{holexp}.
Using the dynamical exponent $z=1/\eta=3/2$ deduced from (4,5) 
one finds that it may be possible to match the conductivity to (39) provided that 
\be
\theta=d-1
\ee
Incidentally, this value of the HV parameter 
was previously singled out on the basis of analyzing entanglement entropy \cite{holferm}.
Besides, it suggests the interpretation of $d-\theta$ as an effective 
number of dimensions orthogonal to the FS. 

The other frequently invoked relation \cite{hol,holferm,hv} is 
\be
z=1+\theta/d
\ee
in which case the first inequality in (27) is marginally satisfied as equality.
Notably, in $2d$ it would only be consistent with (40) for $z=3/2$.

However, these values of the HV exponents imply a sub-linear (electronic) specific heat (38), 
$C(T)\sim T^{2/3}$, contrary to experiment which demonstrates a near-linear - or, at most, 
logarithmically enhanced, $C\sim T\ln T$, dependence \cite{holexp}. 
Since the low-$T$ specific heat is very unlikely to be dominated
by any non-fermionic degrees of freedom, this inconsistency 
could cast additional doubts on the general applicability 
of the HV models to the cuprates. 

Also, from the beginning of the cuprates saga
an even greater fixation has always been on the linear-$T$ dependence  
of resistivity, also observed in a variety of other materials \cite{holexp}. 
Of course, the conductivity scaling with frequency (39) does not readily 
translate into its temperature dependence, as it would be determined by a specific mechanism
of momentum relaxation (i.e., Umklapp, phonons, and/or disorder). 

To this end, the use of the memory matrix technique
yielded a proper conductivity scaling \cite{hol,holexp} in both limits of strong, 
\be
\sigma_s(T)\sim T^{(\theta-2-d)/z}
\ee
as well as weak, 
\be 
\sigma_w(T)\sim T^{2(z-1-\Delta)/z}
\ee
momentum-non-conserving scattering where $\Delta$ is the dimension of the leading translation 
invariance-breaking operator. 
The formulas (42) and (43) agree for $\Delta=z+(d-\theta)/2$ which condition coincides with
that of marginal fulfillment of the 
Harris criterion for the disorder scattering to become a relevant perturbation.  

An alternate interpretation of the linear-$T$ resistivity, $\sigma(T)\sim 1/T$,  
proposed in \cite{hol,holexp} relates it to the FL-like 
entropy, $S(T)\sim C(T)\sim T$.  This school of thought introduces the notion of  
inelastic 'Planckian' scattering rate as  
a potentially single most important scale for thermalization/equilibration/information scrambling (albeit not a rate of momentum relaxation) in strongly interacting systems. 

Viewed under this angle, the naive $T$-dependent counterpart of   
(42) reads $\sigma(T)\sim 1/T^{2/z}S(T)$, with the subsequent 
limits $z\to\infty$, $\theta\to -\infty$, and $-\theta/z\to 1$ taken \cite{hol,holexp}. 
This extreme regime was previously 
discussed in the context of the holographic Gubser-Rocha model \cite{gr} which, 
albeit capable of producing the desired $1/T$ conductivity and 
linear-$T$ specific heat/entropy 
in general agreement with the data on cuprates, also predicts a thermal conductivity 
rising with temperature, $\kappa(T)\sim T^2$ (cf., $\kappa(T)\sim 1/T$ in FL). 
Conceivably, even stronger deviations could be found in the Hall response. 
  
Interestingly, it is the (admittedly, unphysical) model of \cite{phendvk} that so far has  
managed to reproduce a longer list of the power-law dependencies found in the cuprates, 
as compared to the competing schemes \cite{phen}. Unfortunately, such a serendipitous success 
does not offer any immediate insight into the underlying mechanism
of the NFL behavior in the cuprates. 

Furthermore, contrasting the large-r and -$\tau$ asymptotics (31) of the
HV holographic propagators against
their eikonal/bosonization counterparts in search of some agreement 
suggests finite positive values of $\theta$, contrary to the 'Planckian' scenario. 
This observation might further reduce the chances of constructing a consistent HV holographic model of the strange metal phase in the cuprates. 

In part, the deficiencies of the HV-based approach have been
circumvented by the arrival of the 'second SYK wave' \cite{syk2} which 
utilizes the Hamiltonian obtained from the conventional combination 
of a kinetic (quadratic) and interaction (quartic)
terms by randomizing the amplitudes of either one or both of these terms $a$ $la$ $SYK$. 
Making such randomization spatially non-uniform one opens a channel for 
non-conservation of momentum which then gives rise to the linear-$T$ 'Planckian' rate
(on top of a constant). 

Unlike the original SYK model, 
its refined variant manages to deliver a linear-$T$ specific heat 
while discovering that the coefficient in front of the putative 
$\sim\omega^{-2/3}$ term in the optical conductivity vanishes \cite{syk2}.
This is a rather non-trivial finding, as the naive evaluation of the optical conductivity 
$\sigma(\omega)\sim Im (\omega-i\Gamma(\omega))^{-1}=\delta(\omega)+\Gamma(\omega)/\omega^2+\dots$
where the effective (transport) scattering rate $\Gamma(\omega)\sim\Sigma(\omega)(k_{\perp}/k_F)^2\sim
\omega^{4/3}$ at transferred momenta $k_{\perp}\sim\omega^{1/3}$ would have   
yielded the sought-out mid-infrared tail on top of the $\delta$-functional Drude peak. 

In that regard, it might be worth mentioning that after the first 
announcements of reproducing the $\sim\omega^{-2/3}$ behavior holographically 
\cite{2/3} all the later inquiries into this issue reported the lack thereof \cite{no2/3}.
    
Of course, the very existence of different explanations (cf., for example, 
\cite{holexp,phen} and \cite{syk2}) for certain scaling laws 
observed in the cuprates may suggest that their ultimate interpretation is yet to be found. 
It would be, therefore, imperative to strive to extend the list 
of matching properties, akin to \cite{phendvk,phen} as the means of discriminating
between the competing schemes. 

On a related note, it might be worth mentioning the recently proposed pseudo-holographic picture of incipient superconducting pairing in the strange metals \cite{schmalian}.
In this scheme, a doubly charged probe bosonic field which represents the emergent 
pairing order parameter propagates in the background bulk geometry 
dual to the strange-metallic normal state.   
This bulk geometry was found to be the pure $AdS_4$,
thus requiring $\theta=0$, 
in conflict with, both, the asymptotic (39) reproducing the power-law optical conductivity
as well as the 'Planckian' scenario.  

Moreover, the asserted holographic construction of \cite{schmalian} 
suffers from other flaws as well: for one, the role 
of the would-be radial variable is played by the difference $\tau_{12}$ 
between the time arguments of a (generally, non-translationally invariant) bi-local field 
$G_{syk}(\tau_1,\tau_2)$ which, unlike the holographic radius $u$, 
is not positive definite. 
Besides, this purely kinematic construction is independent  
of the interaction strength and, therefore, could be 
equally well applied to a weakly-coupled BCS pairing where no 
meaningful holographic duality would be expected in the first place.\\
  
{\it Stranger Things}\\

\noindent
As per the above discussion, in the sole limit of $N\to 0$ without a concomitant $k_F\to\infty$
one there would be no induced Landau damping term in (1). Instead, the bosonic field mediating the interaction (1) acquires its temporal dispersion via the higher order terms, the most common being a quadratic one,     
$D(\omega,{\bf k})=1/(\omega^2+{\bf k}^2)$. 

In this case the eikonal/bosonization action (10) reads 
\be
S_{eik}(\tau,{\bf r})\sim {|r-i\tau|^2\over {\sqrt {\tau^2+r^2}}}
\ee
The propagator (21) obtained with the use of (44) was shown to develop three distinct FS \cite{bagrov}, thus hinting at  
flattening of the renormalized fermion dispersion and bringing about an intrinsic instability
towards the formation of a 'flat band' characterized by $z>1$ \cite{nosieres,bosrice}.

Interestingly, if the interaction function $D(\omega,{\bf k})$
were to be molded into (1) by choosing $\rho=2$ and $\xi\to -\infty$
the HV exponents would be taking values $z=1,\theta=0$, thus satisfying (41) but not (40). 
On the other hand, matching the holographic 
asymptotics (31) would only be possible for $\theta=2$ which doesn't satisfy (40) either. 

The interactions (1) with $\xi=\rho=1$ also arise in the $2d$ problem 
of half-filled Landau level with unscreened Coulomb interactions \cite{qhe}.
The first-order self-energy (4) now demonstrates a 'marginal FL' behavior 
\be
\Sigma(\omega,k)\sim \omega\ln\omega
\ee
In the $N\to 0$ limit one then obtains the counterpart of (11)
\be
S_{eik}(\tau,{\bf r})\sim {r\over {|r-i\tau|}}\ln|r-i\tau|
\ee
which would be marginally consistent - as far as the overall power-counting 
is concerned - with the holographic action (31) that would yield 
mere constants (or, at most, logarithms) 
for $z=1$ and $\theta=0$ in both large-$r$ and -$\tau$ limits. 

Similar to the discussion of the case $\xi=1$, 
it has been argued that the behavior (46) can only develop in some intermediate 
regime. By contrast, at the longest times 
an essentially free algebraic behavior, $G(\tau, r\to 0)\approx G_0(\tau, r\to 0)$,
was found, whereas at large distances the propagator was predicted to
experience no more than a logarithmic suppression,
$G(\tau\to 0, r)\approx G_0(\tau\to 0, r)/\ln r$ \cite{chubsed}.

Likewise, the 'marginal FL' self-energy (45) and corresponding eikonal
behavior (46) which are suggestive of the 
HV parameters $z=1$ and $\theta=0$ can emerge in all the other situations where  
$\rho=d-1$ while $\xi$ is arbitrary. The list includes the finite density 
$QED_4$ \cite{reizer} and weak-coupling $QCD_4$ \cite{qcd} which, historically, 
offered the first glimpses into the Mother-of-all NFL problem. 

Yet another physically relevant example of the $2d$ interaction (1) with $\rho=2, \xi=0$
corresponds to the so-called 'spin-fermion' model of 
itinerant fermions coupled by antiferromagnetic fluctuations 
where the momentum $\bf k$ is measured with respect to the AFM ordering vector \cite{hertz}. 
Neglecting the (in fact, all-important) complications associated with a 
non-circular shape of the realistic FS one computes (10) in the form 
\be
S_{eik}(\tau,{\bf r})\sim {r\over {|r-i\tau|^{1/2}}}
\ee 
consistent with the self-energy $\Sigma(\omega)\sim\omega^{1/2}$. 
This asymptotic also differs from its would-be holographic counterpart (31) with the HV parameters 
$z=1/\eta=2$ and $\theta=d(z-1)=2$ chosen in accordance with (4) and (41), respectively. 

Switching gears, the continuing interest in graphene and other $2d$ and $3d$ (semi)metals- as well the earlier examples of $1d$ metals (e.g., carbon nanotutes) and  
gapless $2d$ high-$T_c$ superconductors - brought out 
the problem of (pseudo)relativistic Dirac/Weyl fermions with isolated Fermi points
and/or long-range interactions. In those constructions that involve physical (electrically charged) electrons, their $3d$ Coulomb interaction does not get effectively screened, as compared to its bare form, $D(\omega,{\bf k})\sim 1/k^{d-1}$. 

Also, the theory of compressible even-denominator QHE states
that were originally described in terms of the finite density 'composite fermions' \cite{qhe}, 
was later reformulated to elucidate its hidden Dirac properties \cite{son2,qhedirac}.  

In $1d$ the pseudo-relativistic behavior sets in generically in the presence of
partially filled bands. By using the standard bosonization 
one readily obtains the faster-than-algebraic decay of the propagator \cite{schulz}
\be
G_{bos}(\tau, x)\sim \exp(-\ln^{3/2}|x-i\tau|)
\ee
that can be thought of as the $1d$ analog of the Mott insulating state. 
Considering that in $1d$ no Landau damping can occur, the formal values of the HV exponents,
$z=1$ and $\theta=0$, obtained from (4) and (41) for $\rho=0$ and $\xi\to -\infty$ 
yield the asymptotics (31) whose power counting is roughly consistent with (48).  

Lastly, one can mention the putative $U(1)$- 
and $SU(2)$- invariant gauge theories of (pseudo)relativistic neutral and massless fermions (spinons)
in the hypothetical quantum spin liquids \cite{palee}. 
In such $QED_3$-like theories, a sensible gauge-invariant - and potentially 
measurable in, e.g., the ARPES experiments - fermion amplitude
was proposed in the form
\be
G_{inv}(\tau,{\bf r})
=<\psi(\tau,{\bf r})e^{i\int^{(\tau,{\bf r})}_{(0,{\bf 0})}{A}_{\mu}
d{\bf x}^{\mu}}\psi^{\dagger}(0,{\bf 0})>
\ee
which includes a phase factor accumulated along the Wilson line connecting the end points. 
The amplitude (49) was claimed to demonstrate a stronger suppression (faster decay) with 
the Euclidean interval $s={\sqrt {r^2+\tau^2}}$, as compared to the free propagator \cite{dirac}. 

In particular, choosing $\Gamma$ to be the classical straight-line trajectory
was argued to result in the gauge-invariant propagator
$G_{inv}(s)\sim 1/s^{d+\eta_3}$ with a positive anomalous 
dimension $\eta_3\sim 1/N$ \cite{dirac}.
Qualitatively, this power-law behavior implying a logarithmic extremal
action could, once again, be compared to the holographic formulas (31) 
with $\theta=0$, which value would then be consistent with the absence of extended FS.  
 
Contrary to the original expectations, though, the amplitude
(49) was shown to manifest $\eta_3<0$ under the use of a gauge invariance-preserving
regularization scheme \cite{diracdvk}. 

Obviously, being just one (albeit a rather natural one) out of infinitely many other gauge-invariant amplitudes the function (49) may not be necessarily representative of the 
general behavior. In fact, its dependence on  
the contour $\Gamma$ is somewhat reminiscent of that on  
the gauge in the case of the ordinary (non-gauge-invariant, albeit unique) propagator. 

It appears, however, that the important conceptual problem of identifying the proper candidate for
the true physical propagator still remains unsolved. 
Nevertheless, based on the earlier studies of 
massive $QED_4$ \cite{qed}, one would surmise that such an amplitude  
might show a faster-than-algebraic 'super-Luttinger' decay \cite{diracdvk}
\be
G_{inv}(s)\sim \exp(-\ln^2s)
\ee
Also, by analogy with the aforementioned problem of non-relativistic (finite density) fermions
subject to RMF scattering the task of computing physically relevant gauge-invariant fermion amplitudes     
can be addressed in the case of finite temperatures or, equivalently, static (quenched) $QED_3$ as well.
It turns out that the Dirac counterpart of this problem allows for a non-perturbative 
solution, too. It can be used in such physically relevant situations as the properties of normal quasiparticles in the vortex line liquid phase of $d$-wave superconductors \cite{diracstatdvk1}
or electron transport in graphene under the influence of thermal shape fluctuations (in- and out-of-plane phonons) 
\cite{diracstatdvk2}.\\

{\it Epilogue}\\
 
\noindent 
A comparison between the asymptotics of the simplest one-particle amplitudes of fermions
with the physically relevant interaction (1) obtained by several techniques, including such experimental ones as $d>1$ bosonization and 'bottom-up' holography, 
demonstrates a rather mixed level of success. The highest chances of achieving their general agreement (if any) could be found is those situations where the large-distance, short-time behavior is algebraic, requiring a vanishing HV parameter $\theta$ as a necessary condition. In turn, the non-algebraic holographic asymptotics developing for $\theta\neq 0$ tend to be expressly symmetric with respect to exchanging $r$ and $\tau^{1/z}$ while their eikonal/bosonization 
counterparts are not. Furthermore, 
their non-algebraic behavior would be   
restricted to some intermediate distances/times once the FS curvature is taken into account.
 
Importantly, even a perfect match between the holographic and some other (believed to be comparatively better established) results would not provide a 
firm justification for the holographic  
technique itself. Indeed, any results obtained under the assumption of a purely classical (non-dynamical) background metric - which assumption is overwhelmingly common to the practical
applications of the holographic approach - would only pertain to its 'light' version, 
as opposed to the full-fledged one. As to the possible desk-top 
simulations of such a 'holography light' scenario, those have been proposed for several
platforms, including flexible graphene flakes \cite{graphene} and hyperbolic metamaterials \cite{meta}.   
 
Projecting into the future, it seems quite 
likely that the ultimate theory of correlated quantum matter will 
eventually assume a form akin to quantum hydrodynamics formulated in 
terms of the moments of quantum distribution function \cite{holbosdvk}. 
Such a collective-field description of the bulk (a.k.a. 'phase') space 
with the $d$-dimensional momentum providing for the extra dimensions 
could be equally well called either bosonization, or holography.
Regardless of the name, though, taking a full advantage of this formally exact approach 
might turn out to be difficult, especially in the physically relevant cases  
of $N\sim 1$ and moderate coupling strengths. 
 
Nevertheless, there still seems to be no good reason neither for this theory to conform to anything as specific and convenient as the EMD Lagrangian (22), nor for the corresponding holographic 
dictionary to be copy-pasted 'ad verbatim' from string/HEP theory.  

One would hope that exposing the existing controversy over this and related issues 
might be helpful to authors of the future original (of course) 
studies on the topic - as well as their knowledgeable and unbiased (of course) referees. 

This note was compiled, in part, while staying at and being supported by the Aspen Center for Physics 
under the NSF Grant PHY-1607611.



\end{document}